\documentclass[fleqn,usenatbib]{mnras}

\usepackage{newtxtext,newtxmath}

\usepackage[T1]{fontenc}

\usepackage{url}
\usepackage{caption}
\usepackage{threeparttable}  
\usepackage{float}

\DeclareRobustCommand{\VAN}[3]{#2}
\let\VANthebibliography\thebibliography
\def\thebibliography{\DeclareRobustCommand{\VAN}[3]{##3}\VANthebibliography}

\usepackage{graphicx}	
\usepackage{amsmath}	

\title[Estimating Dust SEDs of high-redshift galaxies]{Estimating Dust Temperature and Far-IR Luminosity of High-Redshift Galaxies using ALMA Single-Band Continuum Observations}

\author[Y. Fudamoto et al.]{
Y. Fudamoto,$^{1,2}$\thanks{E-mail: y.fudamoto@aoni.waseda.jp}
A. K. Inoue,$^{1,3}$
Y. Sugahara$^{1,2}$
\\
$^{1}$Waseda Research Institute for Science and Engineering, Faculty of Science and Engineering, Waseda University, 3-4-1 Okubo, Shinjuku, Tokyo 169-8555, Japan\\
$^{2}$National Astronomical Observatory of Japan, 2-21-1, Osawa, Mitaka, Tokyo, Japan\\
$^{3}$Department of Physics, School of Advanced Science and Engineering, Faculty of Science and Engineering, Waseda University,\\ 3-4-1, Okubo, Shinjuku, Tokyo 169-8555, Japan
}

\date{Accepted XXX. Received YYY; in original form ZZZ}

\pubyear{2022}

\begin{document}
\label{firstpage}
\pagerange{\pageref{firstpage}--\pageref{lastpage}}
\maketitle

\begin{abstract}
We present a method that derives the dust temperatures and infrared (IR) luminosities of high-redshift galaxies assuming radiation equilibrium in a simple dust and stellar distribution geometry.
Using public data from the Atacama Large Millimeter/submillimeter Array (ALMA) archive, we studied dust temperatures assuming a clumpy interstellar medium (ISM) model for high-redshift galaxies, then tested the consistency of our results with those obtained using other methods.
We find that a dust distribution model assuming a clumpiness of ${\rm log}\,\xi_{\rm clp}=-1.02\pm0.41$ may accurately represent the ISM of high-redshift star-forming galaxies.
By assuming a value of $\xi_{\rm{clp}}$, our method enables the derivation of dust temperatures and IR luminosities of high-redshift galaxies from dust continuum fluxes and emission sizes obtained from single-band ALMA observations.
to demonstrate the method proposed herein, we determined the dust temperature ($T_{\rm d}=95^{+13}_{-17}\,\rm{K}$) of a $z\sim8.3$ star-forming galaxy, MACS0416-Y1.
Because the method only requires a single-band dust observation to derive a dust temperature, it is more easily accessible than multi-band observations or high-redshift emission line searches and can be applied to large samples of galaxies in future studies using high resolution interferometers such as ALMA.
\end{abstract}

\begin{keywords}
galaxies:high-redshift -- galaxies:ISM -- submillimetre:ISM
\end{keywords}



\section{Introduction} \label{sec:intro}

Investigating star-formation activity in the high-redshift Universe is a key to understanding the origins and evolution of galaxies.
Over the past decades, deep and wide field near-infrared (NIR) surveys have provided large samples of data for high-redshift star-forming galaxies. These surveys have significantly advanced our knowledge about the history of star-formation activity in the Universe, showing the star-formation rate density of the Universe rapidly increases from very high redshift to $z\sim3$, reaches a plateau at $z\sim3-2$, then decreases slowly over $\sim10$ billion years, until $z=0$ (see \citealt{Madau2014} for a review).

The Atacama Large Millimeter/submillimeter Array (ALMA) system has revolutionized our understanding of dust-obscured star-formation activity at high redshift by virtue of its unprecedented sensitivity and angular resolution.
ALMA's follow-up observations of rest-frame UV luminous galaxies have revealed that a large fraction of star-formation activities are dust-obscured, even at high redshift \citep[e.g.,][]{Fudamoto2020,Gruppioni2020,Khusanova2021,Zavala2021,Schouws2022}.
In addition, ALMA has identified the existence of heavily dust-obscured galaxies that are extremely faint; their detection is challenging even when using extremely sensitive rest-frame UV emission observations \citep[e.g.,][]{Wang2019,Fudamoto2021}. These studies have demonstrated the importance of observational constraints on dust-obscured star-formation activities as tools to provide a comprehensive understanding of galaxy growth at high redshift \citep[see][for a review]{Hodge2020}.

However, several studies have noted that estimated infrared luminosities ($L_{\rm IR}$) and, hence, dust-obscured star-formation activities from ALMA observations have large systematic uncertainties \citep[e.g.,][]{Fudamoto2020b,Faisst2020,Sommovigo2021, Ferrara2022}. The major uncertainty comes from the fact that typical ALMA observations provide a very sparse sampling of the rest-frame far-infrared (FIR) spectral energy distributions (SED).
From the limited wavelength coverage of observations, one thus needs to extrapolate the sparsely sampled FIR SED assuming a specific dust temperature to estimate the total $L_{\rm IR}$, considering a wavelength range of $\lambda = 8 - 1000\,{\rm \mu m}$.

Recent efforts to study FIR SEDs of high-redshift galaxies through stacking FIR data over wide wavelength ranges \citep[e.g., ][]{Schreiber2018,Bethermin2020} have revealed that the average dust temperatures of high-redshift galaxies are much higher than those of local galaxies.
Alternatively, although the required observations are very time-consuming, several studies have used small samples of dust temperature measurements combining multi-band ALMA observations to cover a relatively wide wavelength range in the rest-frame FIR \citep{Faisst2020,Bakx2021,Algera2023}.
In addition to direct observational constraints on FIR SEDs, a combination of theoretical models and observed data enabled the prediction of dust temperatures for high-redshift galaxies  \citep[e.g.,][]{Sommovigo2021,Ferrara2022}.

Recently, \citet{Inoue2020} (hereafter, I20) developed a simple alternative analytical approach that enables the derivation of dust temperatures and IR luminosities using single-band dust continuum observations, dust emission sizes, and dust-to-stellar distribution geometry models.
So far, the method is restricted to relatively few examples (I20; \citealt{Sugahara2021}), but has been shown to accurately constrain the dust temperature of high-redshift galaxies.
In this paper, we further examine the applicability of the method developed in I20, calibrate the model using public data from the ALMA archive, and provide Python scripts to derive dust temperature using single ALMA continuum measurements.
The code used herein has been made publicly available\footnote{\url{https://github.com/yfudamoto/FIS22sed.git}} and is readily applicable for deriving dust temperatures and IR luminosities using upcoming ALMA single-band continuum observations.

Herein, we define ``dust temperature'' as luminosity weighted, galaxy-scale averaged temperatures derived from single-temperature modified black body function fits to the observed FIR SEDs of galaxies \citep[e.g.,][]{Casey2012,Liang2019}. Thus, we treat the simplified dust properties of galaxies, as detailed observations that are not available in existing FIR observations of high-redshift galaxies, \citep[such as radial gradients of dust temperature, see e.g.,][for a review]{Galliano2018}.

This paper is organized as follows: in \S2 we describe the samples and observations used. In \S3, we present the methodology applied.  \S4 shows results on the dust temperatures. In \S5, we compare dust temperatures derived using the present approach with those derived using other methods. Finally, we present our conclusions in \S6.
Throughout this paper, we assume a cosmology with $(\Omega_m,\Omega_{\Lambda},h)=(0.3,0.7,0.7)$.

\begin{table*}
	\centering
	\begin{threeparttable}
	\caption{Summary of Galaxy Properties}
      \label{tab:galaxies}
  \begin{tabular}{cccccc}
        \hline\hline
           Name & redshift & UV luminosity & FWHM$_{\rm major}$ &  FWHM$_{\rm minor}$ & Ref.\\ 
	   &    & $\times10^{11}\,{\rm L_{\odot}}$ & kpc &  kpc & \\ 
        \hline
	HZ4 & 5.544 & $1.8\pm0.2$ & $5.4\pm1.2$ & $4.8\pm1.2$ &  1,2,3\\
	HZ6 & 5.293 & $2.1\pm0.1$ & $6.1\pm1.2$ & $4.3\pm1.2$ &  1,2,3\\
	HZ9 & 5.541 & $0.9\pm0.1$ & $1.8\pm0.6$ & $1.8\pm0.6$ &  1,2,3\\
	HZ10 & 5.293 & $2.2\pm0.3$ & $5.5\pm0.6$ & $1.4\pm0.2$ &  1,2,3 \\ 
	A1689\_zD1$^{\ast}$ & 7.13 & $0.21\pm0.02$ & $8.9\pm0.5$ & $2.6\pm0.5$ & 4,5 \\
	B14-65666 & 7.152 & $2.0\pm0.2$ & $3.6\pm1.0$ & $1.0\pm0.5$ & 6,7,8\\
	J0217-0208 & 6.2037 & $4.4\pm0.2$ & $2.8\pm1.1$ & $1.7\pm1.1$ & 9,10 \\
	J1211-0118 & 6.0293 & $4.6\pm0.3$ & $1.3\pm0.2$ & $0.4\pm0.2$ & 9,10\\
	MACS0416-Y1$^{\ast}$ & 8.3118 & $4.1\pm0.1$ & $1.7\pm0.4$ & $0.5\pm0.2$ & 11\\
        \hline
      \end{tabular}
\begin{tablenotes}
\item $\ast$ A16889\_zD1 and MACS0416-Y1 are gravitationally magnified with magnification factors of $\mu=9.3$ and $\mu=1.43$, respectively. Demagnified properties are listed in this table.\\
References: 1: \cite{Capak2015}, 2: \cite{Faisst2020}, 3: \cite{Pavesi2019}, 4: \cite{Watson2015}, 5: \cite{Bakx2021}, 6: \cite{Bowler2018},7: \cite{Hashimoto2022}, 8: \cite{Sugahara2021}, 9: \cite{Matsuoka2018}, 10: \cite{Harikane2020}, 11: \cite{Tamura2019}.
\end{tablenotes}
\end{threeparttable}
\end{table*}

\section{Data}
We used public data from the ALMA archive of nine normal star-forming galaxies
at $z\sim5.5$ to $z\sim8.3$ (see Table \ref{tab:galaxies}).
We selected high-redshift ($z>5$) galaxy observations with relatively high resolution ($<1^{\prime\prime}$) ALMA observations showing significant dust continuum detection in order to assure secure dust emission size measurements. Such secure measurements are required as they comprise an essential parameter in the method developed herein (see \S\ref{sec:method}).
We used galaxies with multiwavelength dust detections in order to calibrate the dust emission model (i.e., HZ4, HZ6, HZ9, HZ10, A1689-zD1, and B14-65666), and applied this model to predict the dust temperatures of galaxies with only single-wavelength continuum detection (i.e., J2017-0208, J1211-0118, and MACS04160-Y1). We briefly describe the galaxies used in this study in the following:

\vspace{0.25cm}
$\bullet$ HZ4 ($z=5.544$), HZ6 ($z=5.293$), HZ9 ($z=5.541$), and HZ10 ($z=5.293$) have ALMA band-6 ($\lambda_{\rm rest}\sim200\,\rm{\mu m}$), band-7 ($\lambda_{\rm rest}\sim160\,\rm{\mu m}$), and band-8 ($\lambda_{\rm rest}\sim110\,\rm{\mu m}$) continuum detections.
Deep multi-band ALMA observations show that the studied galaxies are characterized by robust dust temperature measurements \citep[][]{Pavesi2019,Faisst2020}. Flux density measurements used are from \citet{Faisst2020}.

\vspace{0.25cm}
$\bullet$ J1211-0118 ($z=6.029$) and J0217-0208 ($z=6.204$) have been characterized by multiwavelength ALMA observations using band-6 ($\lambda_{\rm rest}\sim200\,\rm{\mu m}$), band-7 ($\lambda_{\rm rest}\sim160\,\rm{\mu m}$), and band-8 ($\lambda_{\rm rest}\sim90\,\rm{\mu m}$) \citep{Harikane2020}. Although these galaxies have multiwavelength FIR observations, dust temperatures are only poorly constrained due to weak or non-detection in the band-8 continuum, leaving a narrow wavelength coverage of dust continuum detections. Thus, we treat these galaxies as having unconstrained dust temperatures. We used flux density measurements from \citet{Harikane2020}.

\vspace{0.25cm}
$\bullet$ A1689-zD1 ($z=7.13$) is a strongly lensed galaxy with a magnification factor of $\mu=9.3$ \citep{Knudsen2017}. It has been characterized by ALMA observations using band-6 ($\lambda_{\rm rest}\sim160\,\rm{\mu m}$), band-7 ($\lambda_{\rm rest}\sim100\,\rm{\mu m}$), band-8 ($\lambda_{\rm rest}\sim90\,\rm{\mu m}$), and band-9 ($\lambda_{\rm rest}\sim50\,\rm{\mu m}$).  These multiwavelength observations indicate that A1689-zD1 has one of the most robust dust temperature constraints at $z>7$ \citep{Watson2015,Knudsen2017,Bakx2021}. We used flux density measurements from \citet{Bakx2021}, and size measurements from I20.

\vspace{0.25cm}
$\bullet$ B14-65666 ($z=7.15$; also known as {\it Big Three Dragons}) shows two clumps in rest-UV and dust, suggesting an ongoing merger event \citep{Bowler2018,Hashimoto2019,Hashimoto2022}.
B14-65666 has been characterized by multiwavelength ALMA observations using band-6 ($\lambda_{\rm rest}\sim160\,\rm{\mu m}$), band-7 ($\lambda_{\rm rest}\sim120\,\rm{\mu m}$), and band-8 ($\lambda_{\rm rest}\sim90\,\rm{\mu m}$) and is treated as a dust temperature constrained sample.
For the analysis presented herein, we used flux density and size measurements from \citet{Sugahara2021}.

\vspace{0.25cm}
$\bullet$ MACS0416-Y1 ($z=8.3118$) exhibits a significant dust continuum detection in band-8 ($\lambda_{\rm rest}\sim90\,\rm{\mu m}$; \citealt{Tamura2019}); however, observation of the sensitive follow-up band-5 ($\lambda_{\rm rest}\sim160\,\rm{\mu m}$) resulted in non-detection of the dust continuum. The strong lower limits of the dust temperature show extremely high dust temperatures of $T_{\rm d} \gtrsim 80\,\rm{K}$ \citep{Bakx2020}. We therefore treat MACS0416-Y1 as a dust-temperature unconstrained sample. We used flux density measurements from \citet{Bakx2021} and size measurements from \citet{Tamura2019}.

\section{Methods}
\label{sec:method}

\subsection{ISM Geometry and Radiation Model: the Clumpy Sphere Distribution}

We used the dust emission model presented in I20, in which dust absorption and emission assume radiative equilibrium. By incorporating ISM geometry models and the observed sizes of dust emission, I20 successfully developed a simple analytical method by which dust temperature and luminosity could be estimated. In their method, assuming radiative equilibrium and dust geometry models, dust temperature and mass (and thus dust emission luminosity) are not independent variables, but exhibit a one-to-one connection.

I20 considered three different assumptions regarding ISM-to-stellar distributions, i.e., the uniform shell, uniform sphere, and clumpy sphere distributions. Herein, we used the clumpy sphere distribution model with an analytical multi-phase medium \citep[e.g.,][]{Neufeld1991,Hobson1993,Varosi1999,Inoue2005} because it incorporates the complicated geometry of star-forming galaxies at high redshift.

In the clumpy sphere distribution of I20, the ISM and thus the dust, is assumed to reside both in clumps and in interclump media. Clumps and radiation sources are independently and uniformly distributed in a sphere, while the interclump medium fills the rest of the space. Clumps, the interclump medium, and radiation sources are uniformly distributed and have homogeneous properties.
A density contrast is allowed between clumps and the interclump medium.
To simplify the model, I20 further assumed a high-density contrast case, neglecting opacity in the interclump medium, and denoting an extremely tenuous interclump medium. 
In such a case, we can describe the opacity of each clump as $\tau_{\rm{cl}} \approx \tau_{\rm{hom}}\,\xi_{\rm clp}$, where $\tau_{\rm{hom}}$ is the opacity when dust in the system is distributed homogeneously. Here $\xi_{\rm clp}$ is a dimensionless, free parameter controlling the clumpiness of the system, termed the clumpiness parameter.

Under the assumption, I20 describes an ISM and dust geometry model, leaving clumpiness ($\xi_{\rm clp}$) as a free parameter.
In return, assuming $\xi_{\rm clp}$ and measuring the size of the source, the dust geometry can be fully determined such that the radiative equilibrium assumption enables dust SED, and thus temperature and luminosity, to be derived.

Practical use of the I20 model assuming radiative equilibrium and clumpy sphere geometry requires that assumes a $\xi_{\rm cl}$ and uses measurements of rest-frame UV luminosity, redshift, FIR emission size, and FIR flux as observation-based inputs. These measurements are available from moderately high angular resolution ALMA observations, even with single-band dust continuum observations. Therefore, assuming a reasonable value of $\xi_{\rm cl}$, the method is readily applicable and requires only a relatively small amount of observation time compared with observations obtaining multiple data points across a wide wavelength range.

\subsection{Size Measurement}
Although most of the essential physical parameters for the sample used are available from existing literature (see references in Table \ref{tab:galaxies}), dust continuum emission sizes are sometimes unavailable (HZ4, HZ6, HZ9, HZ10, J2017-0208, and J1211-0118). We therefore performed FIR size measurements using significant dust continuum data.

In particular, we performed visibility-based size measurements using the task \path{uv_fit} in the software package GILDAS\footnote{GILDAS is an interferometry data reduction and analysis software developed by Institut de Radioastronomie Millim{\'e}trique (IRAM) and is available from\url{https://www.iram.fr/IRAMFR/GILDAS/}.
To convert ALMA measurement sets to GILDAS/MAPPING uv-tables, we followed the approach described at {\url{ https://www.iram.fr/IRAMFR/ARC/documents/filler/casa-gildas.pdf}}}.
After creating continuum visibility by masking emission lines, we used a single 2D Gaussian source model for data fitting. The free parameters of the fittings are source positions, major/minor axis FWHMs, position angles, and source fluxes.
Size measurements are summarized in Table \ref{tab:galaxies}.

We further evaluated whether the measured FWHMs from fitting visibility data are consistent with those from fitting image plane data. To measure FWHM in the image plane, we created images using the \path{NATURAL} weighting scheme and cleaned these down to three times RMSs, where RMSs are calculated using dirty images. 2D Gaussian fitting using CASA task IMFIT was performed on the cleaned images, showing that the measured sizes match well with those measured from visibility data. For the following analysis, we used FIR sizes obtained from visibility data.

Geometric means were calculated for the best fit major and minor axis FWHMs and used as the radii of the dust continua: $r_{\rm dust} = \sqrt{a\,b}/2$, where $a$ is the major axis FWHM and $b$ is the FWHM of the minor axis. This simplification is required because the I20 method assumes only spherical dust emission geometry. Further studies using more complex dust geometry are beyond the scope of this study.

\begin{figure*}
 \begin{center}
  \includegraphics[width=0.8\textwidth]{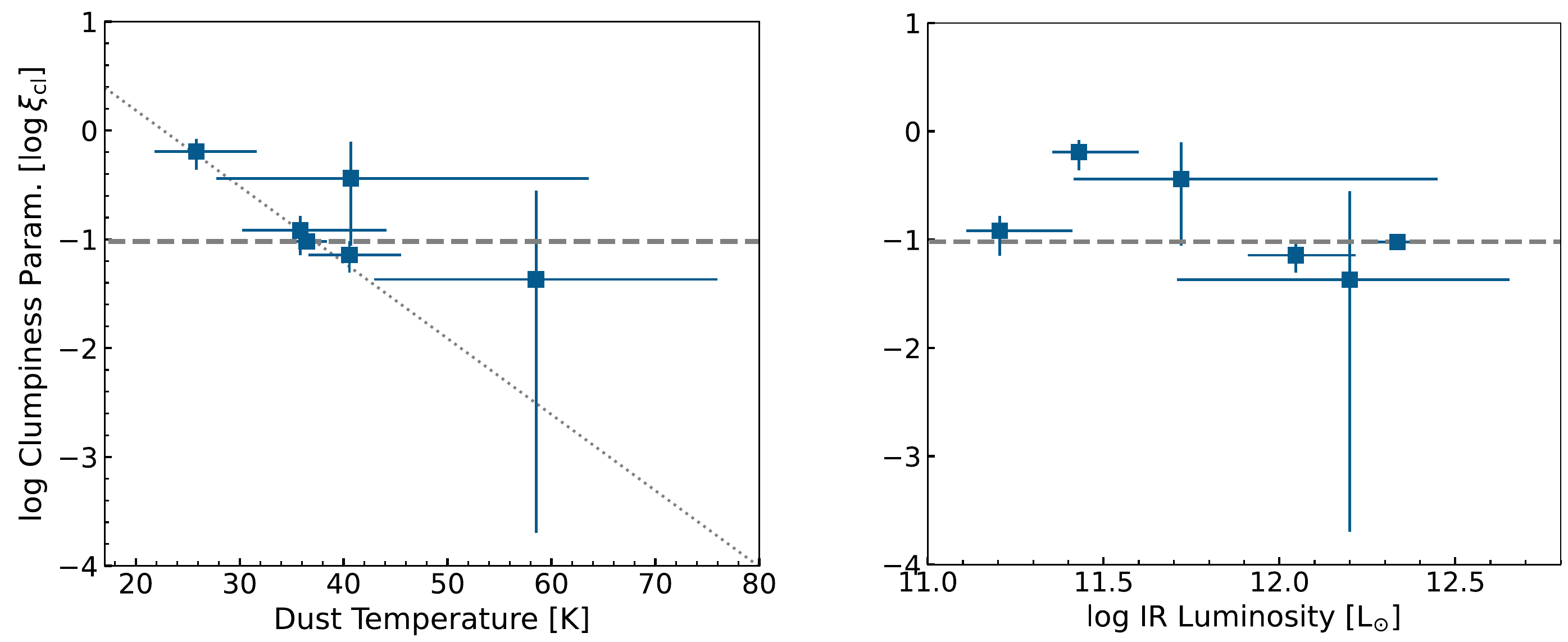} 
 \end{center}
\caption{
Derived clumpiness parameters ($\xi_{\rm clp}$) plotted against dust temperatures (left panel) and IR luminosities (right panel) using the clumpy media dust geometry.
Error bars show $1\,\sigma$ uncertainty estimated using the Monte-Carlo technique.
Our results indicate an average value of ${\rm log}\,\xi_{\rm clp}=-1.0\pm0.4$ (dashed gray line), which we applied to the single-wavelength observation sample. The dotted line in the left panel shows a linear fit to the tentative relation between $\rm{log\,}\xi_{\rm clp}$ and dust temperature.
}\label{fig:Lirclp}
\end{figure*}

\renewcommand{\arraystretch}{1.4}
\begin{table*}
\begin{threeparttable}
 \caption{Summary of the Fitting Results}
 \label{tab:fitresults}
  \begin{tabular}{ccccc}
        \hline\hline
	Name & Dust Temperature & Dust Mass & IR Luminosity & ${\rm log}\,\xi_\mathrm{cl}$ \\ 
	& K & ${\rm log}\,(M_{\rm d}/\rm{M_{\odot}})$ & ${\rm log}\,(L_{\rm d}/\rm{L_{\odot}})$ &  \\ 
         \hline
	HZ4 & $59^{+18}_{-16}$ & $7.1^{+0.3}_{-0.2}$ & $12.2^{+0.5}_{-0.5}$ & $-1.37^{+0.82}_{-2.33}$ \\
	HZ6 & $26^{+6}_{-4}$ & $8.4^{+0.4}_{-0.4}$ & $11.4^{+0.2}_{-0.1}$ & $-0.19^{+0.12}_{-0.17}$ \\
	HZ9 & $41^{+5}_{-4}$ & $7.9^{+0.1}_{-0.1}$ & $12.1^{+0.2}_{-0.1}$ & $-1.14^{+0.13}_{-0.16}$ \\
	HZ10 & $36^{+2}_{-2}$ & $8.5^{+0.1}_{-0.1}$ & $12.3^{+0.1}_{-0.1}$ & $-1.02^{+0.06}_{-0.07}$ \\ 
	A1689-zD1 & $38^{+3}_{-3}$ & $7.2^{+0.1}_{-0.1}$ & $11.3^{+0.1}_{-0.1}$ & $-0.92^{+0.14}_{-0.23}$ \\
	B14-65666 & $41^{+23}_{-13}$ & $7.4^{+0.8}_{-0.5}$ & $11.7^{+0.7}_{-0.3}$ & $-0.44^{+0.34}_{-0.62}$ \\
	\hline\hline
	\multicolumn{5}{c}{fixed $\xi_{\rm clp}$ sample$^{\ast}$}\\
	\hline
	J0217-0208 & $74^{+12}_{-11}$ & $6.6^{+0.2}_{-0.1}$ & $12.4^{+0.3}_{-0.3}$ & --  \\
	J1211-0118 & $59^{+6}_{-8}$ & $7.3^{+0.2}_{-0.1}$ & $12.4^{+0.2}_{-0.1}$ & -- \\
	MACS0416-Y1 & $95^{+13}_{-16}$ & $6.1^{+0.2}_{-0.2}$ & $12.4^{+0.2}_{-0.3}$ & -- \\
        \hline
      \end{tabular}
\begin{tablenotes}
	\item $\ast$: We used ${\rm log}\,\xi_{\rm clp}=-1.02\pm0.41$ for fixed $\xi_{\rm clp}$ objects.
\end{tablenotes}
\end{threeparttable}
\end{table*}

\subsection{Fitting}

For ease of operation, we devised two customized routines using Python: (1) a routine to the estimate dust geometry parameter ($\xi_{\rm clp}$), dust temperature, dust mass, and IR luminosity using multiwavelength FIR continuum measurements (see \S\ref{sec:calibration}), and (2) a routine to estimate the dust temperature, dust mass, and IR luminosity assuming a fixed dust geometry parameter $\xi_{\rm cl}$, using single-wavelength FIR continuum measurements (see \S\ref{sec:application}). All codes and their associated documentation are publicly available (see URL in the \S\ref{sec:intro}).

For the fitting, we used \texttt{minimize} from the Python package \texttt{scipy.optimize} to perform chi-square minimization. Uncertainties in dust temperature, dust mass (IR luminosity), and the clumpiness parameter were estimated using the Monte-Carlo technique by fluctuating the measurements (i.e., UV luminosity, IR size, fluxes of galaxies) 1000 times and assuming Gaussian distributions with measured one sigma errors.
The resulting output distributions were used to estimate the median, 16th, and 84th percentiles, which are treated as the final estimated values and uncertainties, respectively.

Dust emission is assumed to be optically thin modified black body radiation, as is typical used for high-redshift galaxies \citep{Casey2012}, with fixed dust emissivity index ($\beta=2$). The following fiducial dust properties were selected based on a compilation of theoretical, empirical models, and experimental measurements (see Appendix of I20): mass absorption coefficient ($\kappa_{\rm UV}=5.0\times10^4\,{\rm cm^2/g}$) and dust emissivity ($\kappa_{0}=30\,\rm{ cm^2/g}$ at $100\,\rm{\mu m}$). 

\section{Results}

\subsection{Clumpy Medium Model Fitting using Multiwavelength ALMA observations}
\label{sec:calibration}

To calibrate the ISM geometry of the model, we measured the average value of $\xi_{\rm clp}$ for high-redshift galaxies, using the method presented in I20, to galaxies with multiwavelength continuum detections in our sample, specifically HZ4, HZ6, HZ9, HZ10, A1689\_zD1, and B14-65666. These galaxies feature at least three band continuum detections at different wavelengths, covering the rest-frame wavelength of $\lambda_{\rm rest}\sim100\,\rm{\mu m}$  to $\sim200\,\rm{\mu m}$. Thus, they provide an ideal sample for calibrating the clumpiness parameter $\xi_{\rm clp}$ of the ISM distribution geometry model. 
The input parameters are dust emission size, measured flux, UV luminosity, and redshift, and the routine output are dust temperature, dust mass, and ISM clumpiness ($\xi_{\rm clp}$). IR luminosities were calculated using dust mass and dust temperature.
Uncertainties were estimated using the Monte-Carlo technique by fluctuating input parameters assuming Gaussian distributions.
Our results are listed in Table \ref{tab:fitresults} as the free $\xi_{\rm clp}$ sample (see Fig. \ref{fig:SEDs_multi} in the Appendix for the best fit SEDs).

Our fitting accurately reproduced previously obtained dust temperatures and IR luminosities because this and previous studies use almost identical assumptions for dust emission (i.e., optically thin modified blackbody emission). By assuming a clumpy ISM geometry model and using dust emission sizes, we also constrained the ISM clumpiness parameter ($\xi_{\rm clp}$).

We found that the average value of ${\rm log}\,\xi_{\rm clp}=1.02\pm0.41$ (Fig. \ref{fig:Lirclp}) when using an average of ${\rm log}\,\xi_{\rm clp}$ individual galaxies in which the uncertainty shows the standard deviation of individual ${\rm log}\,\xi_{\rm clp}$. 
Although we found no clear dependence of $\xi_{\rm clp}$ on the other properties of galaxies studied, we did identify a very tentative correlation between $\xi_{\rm clp}$ and dust temperature (dotted line in the left panel of Fig. \ref{fig:Lirclp}). We discuss this tentative correlation below in \S\ref{sec:uncertainty}, and 
in the following analysis, we use the typical clumpiness of the model ISM geometry, assuming ${\rm log}\,\xi_{\rm clp}=1.02\pm0.41$.

Systematic uncertainty in the fitting results may arise from our choice of parameters for dust emission. In particular, we selected a dust emissivity index of $\beta = 2$. The dust emissivity index is generally known to degenerate with dust temperatures derived by fitting to modified black body functions (see, e.g., discussions in \citealt{Faisst2020}; I20), although the $\beta$ of each galaxies are poorly constrained and are usually flexibly selected to consider different possible values of $\beta$ \citep[e.g., ][]{Hashimoto2019, Sugahara2021}. We tested the impact of varying $\beta$ in our fitting by considering different values of $\beta$ between $1.5$ to $2.5$. Indeed, although we found systematic changes in the derived median dust temperature and $\xi_{\rm clp}$, these variations are within the $1\,\sigma$ uncertainty derived when fixing $\beta=2.0$ (i.e., typical $\Delta T_{\rm d}\sim5\,\rm{K}$ and $\Delta \rm{log}\,\xi_{\rm d}\sim0.1$). Therefore, the use of a different dust emissivity index does not change our conclusions; rather, it gives minor systematic uncertainty in the derived values. At larger values of $\beta$, the dust temperature becomes lower. Since the $\beta$ values of individual galaxies are poorly constrained, we used the same fixed $\beta$ (i.e., $2$) throughout the calculations.

\subsection{Applying the Clumpy Medium Model using Single-Wavelength ALMA observations}
\label{sec:application}
By assuming the ISM clumpiness obtained in the previous step, we applied our dust emission model to galaxies with only single-band dust continuum detection (i.e., J2017-0208, J1211-0118, and MACS0416-Y1) to estimate their dust temperatures and IR luminosities. 

We used the average parameter, ${\rm log}\,\xi_{\rm clp}=1.02\pm0.41$, estimated in the first step.
We used the same input and output parameters shown in \S\ref{sec:calibration}, but used only input single-band continuum fluxes.
Uncertainties were estimated using the Monte-Carlo technique by fluctuating input parameters assuming Gaussian distributions. Based on this estimation, we obtained reasonable dust temperatures in the range of $51-83\,\rm{K}$. Results are listed as the fixed $\xi_{\rm clp}$ sample in Table \ref{tab:fitresults}. We compare our estimated dust temperature with previous results in \S\ref{sec:comparison}.

\section{Discussion}

\begin{figure}
 \begin{center}
  \includegraphics[width=0.9\columnwidth]{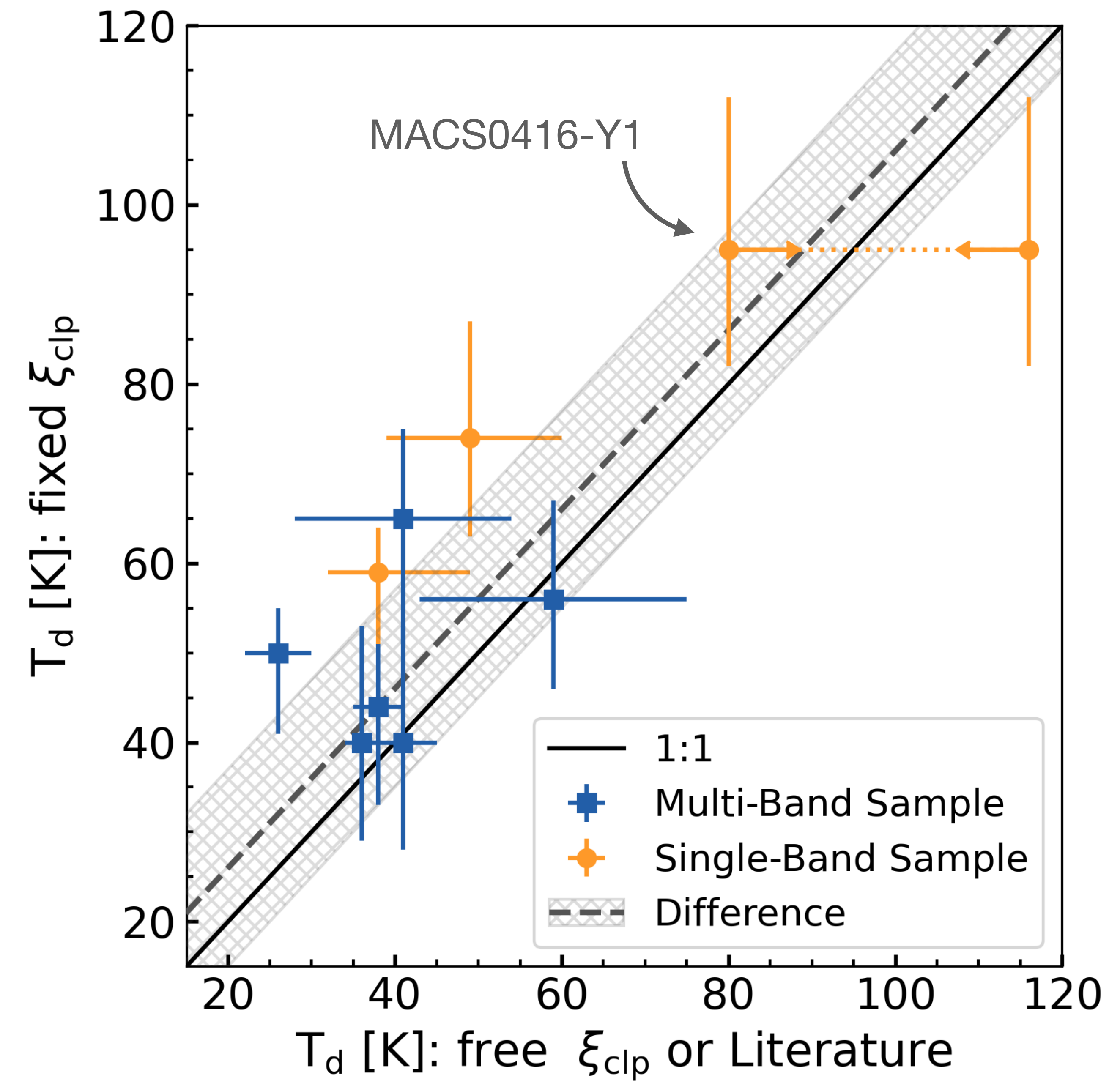} 
 \end{center}
\caption{Comparison of dust temperatures derived using different methods.
Blue squares enable comparison between dust temperatures measured using the free $\xi_{\rm clp}$ multi-band observations (abscissa) and those derived using the fixed $\xi_{\rm clp}$ method assuming ${\rm log}\,\xi_{\rm clp}=-1.02\pm0.41$ (ordinate). 
The distribution of the differences between dust temperatures obtained using different methods are shown with a gray dashed line and a hatched area ($\Delta T=6\pm11\,\rm{K}$).
Yellow circles enable comparison between dust temperatures predicted from existing literature (\citealt{Sommovigo2021, Bakx2021}; abscissa) and those derived using the fixed $\xi_{\rm clp}$ method assuming ${\rm log}\,\xi_{\rm clp}=-1.02\pm0.41$ (ordinate). 
Both methods show consistent dust temperature, at odds of large dispersion suggesting that our method is generally applicable to dust temperature derivation.
}\label{fig:comparison}
\end{figure}

\subsection{Uncertainty of Fixing Dust Geometry}
\label{sec:uncertainty}
Using high-redshift galaxies and assuming ISM geometry, we estimated the dust temperatures and IR luminosities of galaxies even when only single dust emission observations were available (see \S\ref{sec:application}).
However, this approach is possible only when we fix ISM clumpiness ($\xi_{\rm clp}$); furthermore, we need to estimate the uncertainty introduced by fixing this parameter.

To do this, we again used galaxies with multiple dust continuum measurements (i.e., HZ4, HZ6, HZ9, HZ10, A1689\_zD1, and B14-65666). We compared their dust temperatures derived by fitting multiple wavelength data, and those derived assuming ${\rm log}\,\xi_{\rm clp}=-1.02\pm0.41$ using only single dust continuum measurements. For single-band dust temperature estimations, we used the dust continuum measurements with the highest signal to noise ratios.

The results of this treatment are shown in Fig. \ref{fig:comparison}. We found that the dust temperatures derived using both methods generally show one-to-one correspondence; however, there is a relatively large scatter with an average distribution of $\Delta T_{\rm d}=6\pm 11\,\rm{K}$ that is calculated by taking the average and standard deviation of $T_{\rm d}\,({\rm fixed\,\xi_{clp}}) - T_{\rm d}\,({\rm free\,\xi_{clp}})$ (shaded area in Fig. \ref{fig:comparison}). 
This scatter may reflect variations in the dust geometry of individual galaxies (Fig. \ref{fig:Lirclp} and Tab. \ref{tab:fitresults}). Although the comparison shows a relatively large uncertainty, and although a possible systematic overestimation of $T_{\rm d}$ was identified when fixing $\xi_{\rm clp}$, the general agreement of dust temperature suggests that the method works effectively for high-redshift galaxies, especially in the average sense.
Further examination of $\xi_{\rm clp}$ is required using a larger galaxy samples with multiwavelength dust continuum observations and dust emission size measurements in order to study the clumpy ISM geometry model.

Although we simply fixed $\rm{log\,}\xi_{\rm clp}$ to be constant, we also found a tentative correlation between $\xi_{\rm clp}$ and dust temperature (left panel of Fig. 1). This correlation mostly reflects the low dust temperature and high $\xi_{\rm clp}$ of HZ6. Other galaxies show characteristics consistent with the constant $\xi_{\rm clp}$ within an uncertainty of $1\,\sigma$. Thus, due to the lack of statistically significant samples to confirm this relationship, we only tentatively propose this correlation. Nevertheless, we derived this tentative relationship between dust temperatures and $\rm{log\,}\xi_{\rm clp}$ using orthogonal linear regression (the python \texttt{scipy.odr} package) including the uncertainties inherent to both axes. We found that ${\rm Td} = 0.070\,(\pm0.017)\,{\rm log\,}\xi_{\rm clp} + 1.58\,(\pm0.61)$; this relationship entails that galaxies with high dust temperatures have very low $\xi_{\rm clp}$, i.e., ISM geometries similar to that of a homogeneous distribution without any clumps. In contrast, low dust temperature galaxies have high $\xi_{\rm clp}$ and thus more clumpy ISM. confirming these tentative relationships will require larger samples of galaxies.

\subsection{Comparison with other methods}
\label{sec:comparison}

To compare the approach developed herein with existing other methods for deriving dust temperature, we used J0217-0208, J1211-0118, and MACS0416-Y1 because these galaxies have uncertain temperatures or limits of dust temperature only from direct multiwavelength fitting \citep{Harikane2020,Bakx2020}. In particular, we compared with the dust temperatures derived by \citet{Sommovigo2021}, who used [CII] $158\,\rm{\mu m}$ emission line luminosity as an additional parameter for dust temperature and IR luminosity.

The resulting comparisons are shown in Fig. \ref{fig:comparison}. Using the approach developed in this study, we found general agreements with results from existing literature within the uncertainty expected by fixing $\xi_{\rm clp}$. This comparison further strengthens the case that our method provides realistic dust temperature estimations, even when using single-wavelength ALMA observations. In particular, we successfully estimated the dust temperature of MACS0416-Y1 for the first time. In previous studies, MACS0416-Y1 was given only a lower limit \citep{Bakx2020} and upper limit \citep{Sommovigo2021} for its dust temperature. Our analysis of MACS0416-Y1 shows dust temperatures consistent with both its upper and lower limit estimates (Fig. \ref{fig:comparison}).

\subsection{Future Prospects}
Although several caveats still exist in our method, it has several advantage for application to high-redshift galaxy observations. In particular, our method requires only marginal resolution observations of the dust continuum to estimate sizes and fluxes, and therefore requires much shorter observational efforts than scanning emission lines, which are sensitive to gas masses. In future work, we suggest that the relationships between ISM clumpiness, dust emission size, and the dust temperature of lower redshift galaxies (e.g., at $z<4$) should be studied.

Our method can be applied to statistically large amounts of high-redshift galaxies; these data will soon become available after the deep and high resolution survey by the James Webb Space Telescope and the expected follow-up observations by ALMA.

\section{Conclusion}
In this study, we developed a radiation equilibrium method to estimate dust temperatures in high-redshift galaxies, based on the method presented by \citet{Inoue2020}, and applied this method to existing ALMA observations.

We first calibrated the ISM clumpiness parameter in the dust and source distribution geometry model using galaxies with multiwavelength dust continuum measurements, finding an average clumpiness parameter of ${\rm log}\,\xi_{\rm clp}=1.02\pm0.41$.

We then applied the method to galaxies for which only a single-band detection has been made. These single-band detections enabled a robust determination of dust continuum emission size, but no constraints could be placed on dust temperature.
Combining dust continuum size measurements and a simple FIR radiation transfer model, our method successfully measured dust temperatures using single dust continuum detections.

In comparison with other methods, such as multiwavelength fittings \citep[e.g.,][]{Bakx2020} and the [CII] line-based method introduced in \citet{Sommovigo2021}, our method, fixing ${\rm log}\,\xi_{\rm clp}=1.02\pm0.41$ was found to accurately estimate the dust temperatures of high-redshift galaxies; this was confirmed by demonstrating general agreements with several established methods. 

Further constraining and evaluating dust geometry models requires relatively high resolution ALMA observations (i.e., beam sizes of $<0.5^{\prime\prime}$) for high-redshift galaxies with multiwavelength dust continuum measurements. These observations are particularly important for evaluating and reducing systematic uncertainty introduced by assuming fixed $\xi_{\rm clp}$.
Nevertheless, our method can readily estimate the dust temperatures and IR luminosities of high-redshift galaxies for which only have existing and future ALMA single-band dust continuum measurement exist.

\section*{Acknowledgements}

\noindent This paper makes use of the following ALMA data: 
\path{ADS/JAO.ALMA#2018.1.00348.S},
\path{ADS/JAO.ALMA#2017.1.00508.S},
\path{ADS/JAO.ALMA#2011.1.00319.S},
\path{ADS/JAO.ALMA#2012.1.00216.S},
\path{ADS/JAO.ALMA#2013.1.01064.S},
\path{ADS/JAO.ALMA#2016.1.00954.S},
\path{ADS/JAO.ALMA#2019.1.01778.S},
\path{ADS/JAO.ALMA# 2017.1.00225.S},
\path{ADS/JAO.ALMA#2016.1.00117.S},
\path{ADS/JAO.ALMA#2013.1.01064.S},
\path{ADS/JAO.ALMA##2019.1.01491.S},
\path{ADS/JAO.ALMA##2015.1.00540.S},
\path{ADS/JAO.ALMA#2016.1.00954.S},
\path{ADS/JAO.ALMA#2017.1.00190.S}. 
ALMA is a partnership of ESO (representing its member states), NSF(USA) and NINS (Japan), together with NRC (Canada), MOST and ASIAA (Taiwan), and KASI (Republic of Korea), in cooperation with the Republic of Chile. The Joint ALMA Observatory is operated by ESO, AUI/NRAO and NAOJ.
YF, YS, and AKI acknowledge support from NAOJ ALMA Scientific Research Grant number 2020-16B.

\section*{Data Availability}

The ALMA data presented in this paper are publicly available via
the ALMA archive\footnote{\url{https://almascience.nrao.edu/aq/}}.
The python script used in this work is publicly distributed \footnote{\url{https://github.com/yfudamoto/FIS22sed.git}}.



\bibliographystyle{mnras}
\bibliography{base} 

\begin{thebibliography}{}
\makeatletter
\relax
\def\mn@urlcharsother{\let\do\@makeother \do\$\do\&\do\#\do\^\do\_\do\%\do\~}
\def\mn@doi{\begingroup\mn@urlcharsother \@ifnextchar [ {\mn@doi@}
  {\mn@doi@[]}}
\def\mn@doi@[#1]#2{\def\@tempa{#1}\ifx\@tempa\@empty \href
  {http://dx.doi.org/#2} {doi:#2}\else \href {http://dx.doi.org/#2} {#1}\fi
  \endgroup}
\def\mn@eprint#1#2{\mn@eprint@#1:#2::\@nil}
\def\mn@eprint@arXiv#1{\href {http://arxiv.org/abs/#1} {{\tt arXiv:#1}}}
\def\mn@eprint@dblp#1{\href {http://dblp.uni-trier.de/rec/bibtex/#1.xml}
  {dblp:#1}}
\def\mn@eprint@#1:#2:#3:#4\@nil{\def\@tempa {#1}\def\@tempb {#2}\def\@tempc
  {#3}\ifx \@tempc \@empty \let \@tempc \@tempb \let \@tempb \@tempa \fi \ifx
  \@tempb \@empty \def\@tempb {arXiv}\fi \@ifundefined
  {mn@eprint@\@tempb}{\@tempb:\@tempc}{\expandafter \expandafter \csname
  mn@eprint@\@tempb\endcsname \expandafter{\@tempc}}}

\bibitem[\protect\citeauthoryear{{Algera} et~al.,}{{Algera}
  et~al.}{2023}]{Algera2023}
{Algera} H.,  et~al., 2023, \mn@doi [arXiv e-prints]
  {10.48550/arXiv.2301.09659}, \href
  {https://ui.adsabs.harvard.edu/abs/2023arXiv230109659A} {p. arXiv:2301.09659}

\bibitem[\protect\citeauthoryear{{Bakx} et~al.,}{{Bakx}
  et~al.}{2020}]{Bakx2020}
{Bakx} T. J.~L.~C.,  et~al., 2020, \mn@doi [\mnras] {10.1093/mnras/staa509},
  \href {https://ui.adsabs.harvard.edu/abs/2020MNRAS.493.4294B} {493, 4294}

\bibitem[\protect\citeauthoryear{{Bakx} et~al.,}{{Bakx}
  et~al.}{2021}]{Bakx2021}
{Bakx} T. J.~L.~C.,  et~al., 2021, \mn@doi [\mnras] {10.1093/mnrasl/slab104},
  \href {https://ui.adsabs.harvard.edu/abs/2021MNRAS.508L..58B} {508, L58}

\bibitem[\protect\citeauthoryear{{B{\'e}thermin} et~al.,}{{B{\'e}thermin}
  et~al.}{2020}]{Bethermin2020}
{B{\'e}thermin} M.,  et~al., 2020, \mn@doi [\aap]
  {10.1051/0004-6361/202037649}, \href
  {https://ui.adsabs.harvard.edu/abs/2020A&A...643A...2B} {643, A2}

\bibitem[\protect\citeauthoryear{{Bowler}, {Bourne}, {Dunlop}, {McLure}  \&
  {McLeod}}{{Bowler} et~al.}{2018}]{Bowler2018}
{Bowler} R.~A.~A.,  {Bourne} N.,  {Dunlop} J.~S.,  {McLure} R.~J.,   {McLeod}
  D.~J.,  2018, \mn@doi [\mnras] {10.1093/mnras/sty2368}, \href
  {https://ui.adsabs.harvard.edu/abs/2018MNRAS.481.1631B} {481, 1631}

\bibitem[\protect\citeauthoryear{{Capak} et~al.,}{{Capak}
  et~al.}{2015}]{Capak2015}
{Capak} P.~L.,  et~al., 2015, \mn@doi [\nat] {10.1038/nature14500}, \href
  {https://ui.adsabs.harvard.edu/abs/2015Natur.522..455C} {522, 455}

\bibitem[\protect\citeauthoryear{{Casey}}{{Casey}}{2012}]{Casey2012}
{Casey} C.~M.,  2012, \mn@doi [\mnras] {10.1111/j.1365-2966.2012.21455.x},
  \href {https://ui.adsabs.harvard.edu/abs/2012MNRAS.425.3094C} {425, 3094}

\bibitem[\protect\citeauthoryear{{Faisst} et~al.,}{{Faisst}
  et~al.}{2020}]{Faisst2020}
{Faisst} A.~L.,  et~al., 2020, \mn@doi [\apjs] {10.3847/1538-4365/ab7ccd},
  \href {https://ui.adsabs.harvard.edu/abs/2020ApJS..247...61F} {247, 61}

\bibitem[\protect\citeauthoryear{{Ferrara} et~al.,}{{Ferrara}
  et~al.}{2022}]{Ferrara2022}
{Ferrara} A.,  et~al., 2022, \mn@doi [\mnras] {10.1093/mnras/stac460}, \href
  {https://ui.adsabs.harvard.edu/abs/2022MNRAS.512...58F} {512, 58}

\bibitem[\protect\citeauthoryear{{Fudamoto} et~al.,}{{Fudamoto}
  et~al.}{2020a}]{Fudamoto2020b}
{Fudamoto} Y.,  et~al., 2020a, \mn@doi [\mnras] {10.1093/mnras/stz3248}, \href
  {https://ui.adsabs.harvard.edu/abs/2020MNRAS.491.4724F} {491, 4724}

\bibitem[\protect\citeauthoryear{{Fudamoto} et~al.,}{{Fudamoto}
  et~al.}{2020b}]{Fudamoto2020}
{Fudamoto} Y.,  et~al., 2020b, \mn@doi [\aap] {10.1051/0004-6361/202038163},
  \href {https://ui.adsabs.harvard.edu/abs/2020A&A...643A...4F} {643, A4}

\bibitem[\protect\citeauthoryear{{Fudamoto} et~al.,}{{Fudamoto}
  et~al.}{2021}]{Fudamoto2021}
{Fudamoto} Y.,  et~al., 2021, \mn@doi [\nat] {10.1038/s41586-021-03846-z},
  \href {https://ui.adsabs.harvard.edu/abs/2021Natur.597..489F} {597, 489}

\bibitem[\protect\citeauthoryear{{Galliano}, {Galametz}  \& {Jones}}{{Galliano}
  et~al.}{2018}]{Galliano2018}
{Galliano} F.,  {Galametz} M.,   {Jones} A.~P.,  2018, \mn@doi [\araa]
  {10.1146/annurev-astro-081817-051900}, \href
  {https://ui.adsabs.harvard.edu/abs/2018ARA&A..56..673G} {56, 673}

\bibitem[\protect\citeauthoryear{{Gruppioni} et~al.,}{{Gruppioni}
  et~al.}{2020}]{Gruppioni2020}
{Gruppioni} C.,  et~al., 2020, \mn@doi [\aap] {10.1051/0004-6361/202038487},
  \href {https://ui.adsabs.harvard.edu/abs/2020A&A...643A...8G} {643, A8}

\bibitem[\protect\citeauthoryear{{Harikane} et~al.,}{{Harikane}
  et~al.}{2020}]{Harikane2020}
{Harikane} Y.,  et~al., 2020, \mn@doi [\apj] {10.3847/1538-4357/ab94bd}, \href
  {https://ui.adsabs.harvard.edu/abs/2020ApJ...896...93H} {896, 93}

\bibitem[\protect\citeauthoryear{{Hashimoto} et~al.,}{{Hashimoto}
  et~al.}{2019}]{Hashimoto2019}
{Hashimoto} T.,  et~al., 2019, \mn@doi [\pasj] {10.1093/pasj/psz049}, \href
  {https://ui.adsabs.harvard.edu/abs/2019PASJ...71...71H} {71, 71}

\bibitem[\protect\citeauthoryear{{Hashimoto} et~al.,}{{Hashimoto}
  et~al.}{2022}]{Hashimoto2022}
{Hashimoto} T.,  et~al., 2022, arXiv e-prints, \href
  {https://ui.adsabs.harvard.edu/abs/2022arXiv220301345H} {p. arXiv:2203.01345}

\bibitem[\protect\citeauthoryear{{Hobson} \& {Padman}}{{Hobson} \&
  {Padman}}{1993}]{Hobson1993}
{Hobson} M.~P.,  {Padman} R.,  1993, \mn@doi [\mnras]
  {10.1093/mnras/264.1.161}, \href
  {https://ui.adsabs.harvard.edu/abs/1993MNRAS.264..161H} {264, 161}

\bibitem[\protect\citeauthoryear{{Hodge} \& {da Cunha}}{{Hodge} \& {da
  Cunha}}{2020}]{Hodge2020}
{Hodge} J.~A.,  {da Cunha} E.,  2020, \mn@doi [Royal Society Open Science]
  {10.1098/rsos.200556}, \href
  {https://ui.adsabs.harvard.edu/abs/2020RSOS....700556H} {7, 200556}

\bibitem[\protect\citeauthoryear{{Inoue}}{{Inoue}}{2005}]{Inoue2005}
{Inoue} A.~K.,  2005, \mn@doi [\mnras] {10.1111/j.1365-2966.2005.08890.x},
  \href {https://ui.adsabs.harvard.edu/abs/2005MNRAS.359..171I} {359, 171}

\bibitem[\protect\citeauthoryear{{Inoue}, {Hashimoto}, {Chihara}  \&
  {Koike}}{{Inoue} et~al.}{2020}]{Inoue2020}
{Inoue} A.~K.,  {Hashimoto} T.,  {Chihara} H.,   {Koike} C.,  2020, \mn@doi
  [\mnras] {10.1093/mnras/staa1203}, \href
  {https://ui.adsabs.harvard.edu/abs/2020MNRAS.495.1577I} {495, 1577}

\bibitem[\protect\citeauthoryear{{Khusanova} et~al.,}{{Khusanova}
  et~al.}{2021}]{Khusanova2021}
{Khusanova} Y.,  et~al., 2021, \mn@doi [\aap] {10.1051/0004-6361/202038944},
  \href {https://ui.adsabs.harvard.edu/abs/2021A&A...649A.152K} {649, A152}

\bibitem[\protect\citeauthoryear{{Knudsen}, {Watson}, {Frayer}, {Christensen},
  {Gallazzi}, {Micha{\l}owski}, {Richard}  \& {Zavala}}{{Knudsen}
  et~al.}{2017}]{Knudsen2017}
{Knudsen} K.~K.,  {Watson} D.,  {Frayer} D.,  {Christensen} L.,  {Gallazzi} A.,
   {Micha{\l}owski} M.~J.,  {Richard} J.,   {Zavala} J.,  2017, \mn@doi
  [\mnras] {10.1093/mnras/stw3066}, \href
  {https://ui.adsabs.harvard.edu/abs/2017MNRAS.466..138K} {466, 138}

\bibitem[\protect\citeauthoryear{{Liang} et~al.,}{{Liang}
  et~al.}{2019}]{Liang2019}
{Liang} L.,  et~al., 2019, \mn@doi [\mnras] {10.1093/mnras/stz2134}, \href
  {https://ui.adsabs.harvard.edu/abs/2019MNRAS.489.1397L} {489, 1397}

\bibitem[\protect\citeauthoryear{{Madau} \& {Dickinson}}{{Madau} \&
  {Dickinson}}{2014}]{Madau2014}
{Madau} P.,  {Dickinson} M.,  2014, \mn@doi [\araa]
  {10.1146/annurev-astro-081811-125615}, \href
  {https://ui.adsabs.harvard.edu/abs/2014ARA&A..52..415M} {52, 415}

\bibitem[\protect\citeauthoryear{{Matsuoka} et~al.,}{{Matsuoka}
  et~al.}{2018}]{Matsuoka2018}
{Matsuoka} Y.,  et~al., 2018, \mn@doi [\pasj] {10.1093/pasj/psx046}, \href
  {https://ui.adsabs.harvard.edu/abs/2018PASJ...70S..35M} {70, S35}

\bibitem[\protect\citeauthoryear{{Neufeld}}{{Neufeld}}{1991}]{Neufeld1991}
{Neufeld} D.~A.,  1991, \mn@doi [\apjl] {10.1086/185983}, \href
  {https://ui.adsabs.harvard.edu/abs/1991ApJ...370L..85N} {370, L85}

\bibitem[\protect\citeauthoryear{{Pavesi}, {Riechers}, {Faisst}, {Stacey}  \&
  {Capak}}{{Pavesi} et~al.}{2019}]{Pavesi2019}
{Pavesi} R.,  {Riechers} D.~A.,  {Faisst} A.~L.,  {Stacey} G.~J.,   {Capak}
  P.~L.,  2019, \mn@doi [\apj] {10.3847/1538-4357/ab3a46}, \href
  {https://ui.adsabs.harvard.edu/abs/2019ApJ...882..168P} {882, 168}

\bibitem[\protect\citeauthoryear{{Schouws} et~al.,}{{Schouws}
  et~al.}{2022}]{Schouws2022}
{Schouws} S.,  et~al., 2022, \mn@doi [\apj] {10.3847/1538-4357/ac4605}, \href
  {https://ui.adsabs.harvard.edu/abs/2022ApJ...928...31S} {928, 31}

\bibitem[\protect\citeauthoryear{{Schreiber}, {Elbaz}, {Pannella}, {Ciesla},
  {Wang}  \& {Franco}}{{Schreiber} et~al.}{2018}]{Schreiber2018}
{Schreiber} C.,  {Elbaz} D.,  {Pannella} M.,  {Ciesla} L.,  {Wang} T.,
  {Franco} M.,  2018, \mn@doi [\aap] {10.1051/0004-6361/201731506}, \href
  {https://ui.adsabs.harvard.edu/abs/2018A&A...609A..30S} {609, A30}

\bibitem[\protect\citeauthoryear{{Sommovigo}, {Ferrara}, {Carniani}, {Zanella},
  {Pallottini}, {Gallerani}  \& {Vallini}}{{Sommovigo}
  et~al.}{2021}]{Sommovigo2021}
{Sommovigo} L.,  {Ferrara} A.,  {Carniani} S.,  {Zanella} A.,  {Pallottini} A.,
   {Gallerani} S.,   {Vallini} L.,  2021, \mn@doi [\mnras]
  {10.1093/mnras/stab720}, \href
  {https://ui.adsabs.harvard.edu/abs/2021MNRAS.503.4878S} {503, 4878}

\bibitem[\protect\citeauthoryear{{Sugahara} et~al.,}{{Sugahara}
  et~al.}{2021}]{Sugahara2021}
{Sugahara} Y.,  et~al., 2021, \mn@doi [\apj] {10.3847/1538-4357/ac2a36}, \href
  {https://ui.adsabs.harvard.edu/abs/2021ApJ...923....5S} {923, 5}

\bibitem[\protect\citeauthoryear{{Tamura} et~al.,}{{Tamura}
  et~al.}{2019}]{Tamura2019}
{Tamura} Y.,  et~al., 2019, \mn@doi [\apj] {10.3847/1538-4357/ab0374}, \href
  {https://ui.adsabs.harvard.edu/abs/2019ApJ...874...27T} {874, 27}

\bibitem[\protect\citeauthoryear{{V{\'a}rosi} \& {Dwek}}{{V{\'a}rosi} \&
  {Dwek}}{1999}]{Varosi1999}
{V{\'a}rosi} F.,  {Dwek} E.,  1999, \mn@doi [\apj] {10.1086/307729}, \href
  {https://ui.adsabs.harvard.edu/abs/1999ApJ...523..265V} {523, 265}

\bibitem[\protect\citeauthoryear{{Wang} et~al.,}{{Wang}
  et~al.}{2019}]{Wang2019}
{Wang} T.,  et~al., 2019, \mn@doi [\nat] {10.1038/s41586-019-1452-4}, \href
  {https://ui.adsabs.harvard.edu/abs/2019Natur.572..211W} {572, 211}

\bibitem[\protect\citeauthoryear{{Watson}, {Christensen}, {Knudsen}, {Richard},
  {Gallazzi}  \& {Micha{\l}owski}}{{Watson} et~al.}{2015}]{Watson2015}
{Watson} D.,  {Christensen} L.,  {Knudsen} K.~K.,  {Richard} J.,  {Gallazzi}
  A.,   {Micha{\l}owski} M.~J.,  2015, \mn@doi [\nat] {10.1038/nature14164},
  \href {https://ui.adsabs.harvard.edu/abs/2015Natur.519..327W} {519, 327}

\bibitem[\protect\citeauthoryear{{Zavala} et~al.,}{{Zavala}
  et~al.}{2021}]{Zavala2021}
{Zavala} J.~A.,  et~al., 2021, \mn@doi [\apj] {10.3847/1538-4357/abdb27}, \href
  {https://ui.adsabs.harvard.edu/abs/2021ApJ...909..165Z} {909, 165}

\makeatother
\end{thebibliography}




\appendix

\section{SED fits to individual galaxies}

\begin{figure*}[h]
 \begin{center}
  \includegraphics[width=0.9\textwidth]{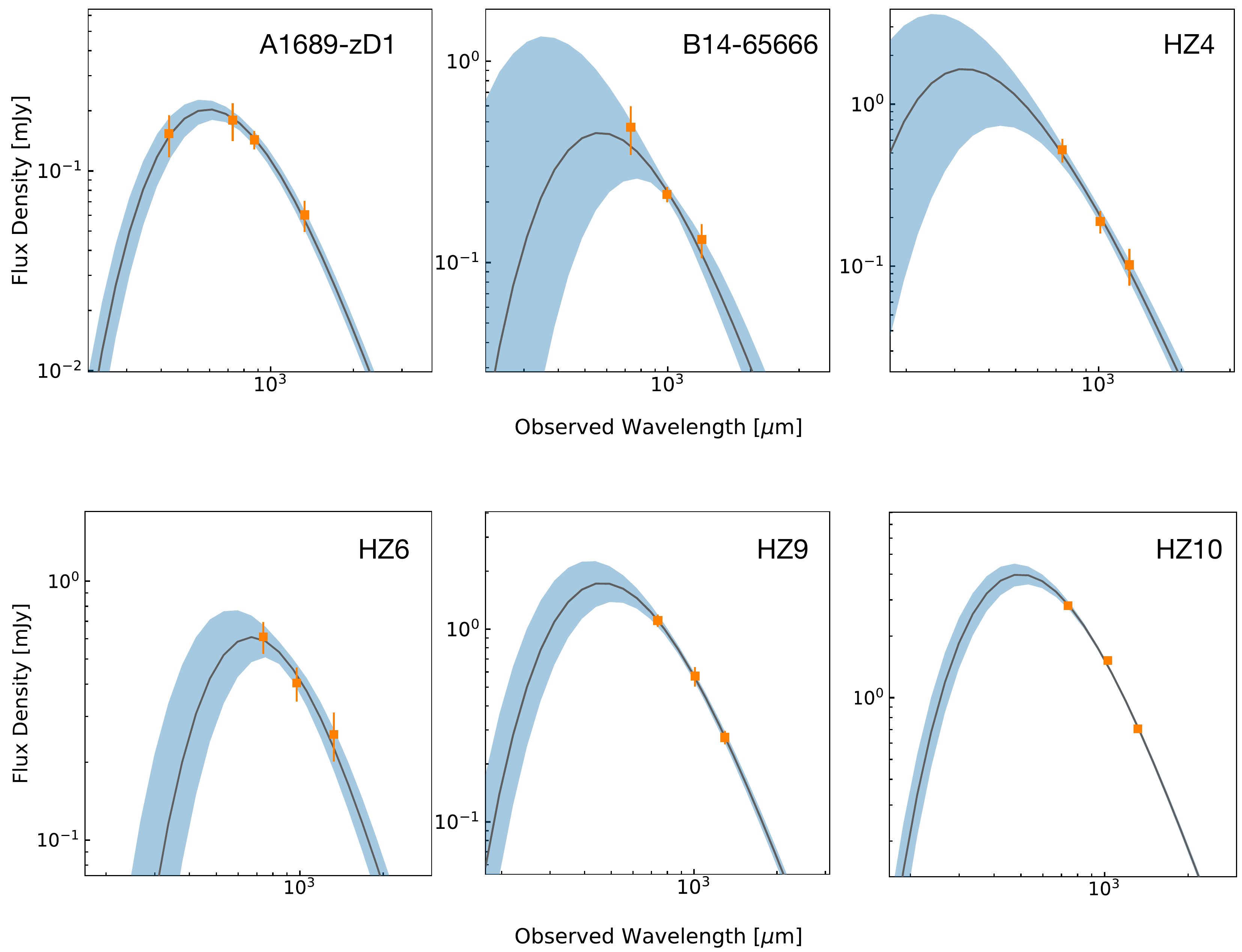} 
 \end{center}
\caption{Best fit FIR SEDs of galaxies for which multiwavelength observations have been acquired using the $\xi_{\rm clp}$-free clumpy ISM model (\S\ref{sec:calibration}). Solid lines denote median values and bands show the 16th to 84th percentiles of the resulting distributions. Outputs of the fittings are summarized in Tab. \ref{tab:fitresults}.
}\label{fig:SEDs_multi}
\end{figure*}

\bsp	
\label{lastpage}
\end{document}